\documentclass[aps,prd,preprintnumbers,superscriptaddress,nofootinbib,floatfix,twocolumn,notitlepage]{revtex4-2}
\usepackage{paralist}
\usepackage{amsmath}
\usepackage{geometry}

\usepackage{dcolumn}   
\usepackage{amssymb,mathrsfs}\usepackage{hyperref}
\hypersetup{%
    colorlinks  = true,%
    linkcolor   = [rgb]{0,0.3,0.7},%
    citecolor   = [rgb]{0,0.3,0.7},%
    urlcolor    = [rgb]{0,0.3,0.7},%
    linktocpage = true%
}
\usepackage{gensymb}
\usepackage{slashed}
\usepackage{graphicx,epsfig,dcolumn,bm,xspace}
\usepackage{booktabs}

\begin{document}
\preprint{KEK-TH-2587}

\renewcommand{\Im}{{\cal I}m}
\renewcommand{\Re}{{\cal R}e}

\newcommand{\fig}[2]{\begin{figure}[H]\begin{center}\includegraphics[height={#1}]{{#2}}\end{center}\end{figure}}

\newcommand{\del}{\partial}
\newcommand{\herm}{\dagger}
\newcommand{\D}{\mathrm{d}}
\newcommand{\lsim}{\raisebox{-0.13cm}{~\shortstack{$<$ \\[-0.07cm]
      $\sim$}}~}
      
\newcommand{\ve}[1]{\ensuremath{\vec{\bf{#1}}}\xspace}
\newcommand{\p}{\ve{p}}

\newcommand{\ket}[1]{|{#1} \rangle}
\newcommand{\bra}[1]{\langle {#1}|}
\newcommand{\mean}[1]{\left\langle {#1}\right\rangle}
\newcommand{\bk}[2]{\langle{#1}|{#2}\rangle}
\newcommand{\mel}[3]{\langle{#2}|{#1}|{#3}\rangle}

\newcommand{\gmu}{\gamma^{\mu}}
\newcommand{\gnu}{\gamma^{\nu}}
\newcommand{\gsig}{\gamma^{\sigma}}
\newcommand{\grho}{\gamma^{\rho}}
\newcommand{\gMu}{\gamma_{\mu}}
\newcommand{\gNu}{\gamma_{\nu}}
\newcommand{\gSig}{\gamma_{\sigma}}
\newcommand{\gRho}{\gamma_{\rho}}
\newcommand{\gfive}{\gamma_5}

\newcommand{\uspinor}[2]{u_{#1}(\vec{#2})}
\newcommand{\buspinor}[2]{\bar{u}_{#1}(\vec{#2})}
\newcommand{\vspinor}[2]{v_{#1}(\vec{#2})}
\newcommand{\bvspinor}[2]{\bar{v}_{#1}(\vec{#2})}

\newcommand{\qrecsq}{q_{\rm rec}^2}
\newcommand{\qrec}{q_{\rm rec}}



\def\beq{\begin{equation}}
\def\eeq{\end{equation}}

\title{Decoding the $B \to K \nu \nu$ excess at Belle~II: kinematics, operators, and masses}

\author{K\r{a}re Fridell}
\email{kare.fridell@kek.jp}
\affiliation{KEK Theory Center, Tsukuba, Ibaraki 305--0801, Japan}
\affiliation{Department of Physics, Florida State University, Tallahassee, FL 32306-4350, USA}

\author{Mitrajyoti Ghosh}
\email{mghosh2@fsu.edu}
\affiliation{Department of Physics, Florida State University, Tallahassee, FL 32306-4350, USA}

\author{Takemichi Okui}
\email{tokui@fsu.edu}
\affiliation{Department of Physics, Florida State University, Tallahassee, FL 32306-4350, USA}
\affiliation{KEK Theory Center, Tsukuba, Ibaraki 305--0801, Japan}

\author{Kohsaku Tobioka}
\email{ktobioka@fsu.edu}
\affiliation{Department of Physics, Florida State University, Tallahassee, FL 32306-4350, USA}
\affiliation{KEK Theory Center, Tsukuba, Ibaraki 305--0801, Japan}

\begin{abstract}
\noindent An excess in the branching fraction for $B^+ \to K^+ \nu\nu$ recently measured at Belle II may be a hint of new physics. We perform thorough likelihood analyses for different new physics scenarios such as $B \to KX$ with a new invisible particle $X$, or $B\to K\chi\chi$ through a scalar, vector, or tensor current with $\chi$ being a new invisible particle or a neutrino. We find that vector-current 3-body decay with $m_X \simeq 0.6$~GeV---which may be dark matter---is most favored, while 2-body decay with $m_X \simeq 2$~GeV is also competitive. The best-fit branching fractions for the scalar and tensor cases are a few times larger than for the 2-body and vector cases. Past BaBar measurements provide further discrimination, although the best-fit parameters stay similar.  
\end{abstract}

\maketitle


\section{Introduction} 
\label{sec:intro}
The decay $B^+ \to K^+ \nu\nu$ is a rare process in the Standard Model (SM)\@. It has long been a sought-after experimental target~\cite{Belle:2013tnz, Belle:2017oht, BaBar:2010oqg, BaBar:2013npw} as it is one of the cleanest channels for the search for new physics owing to well-controlled theoretical uncertainties in the SM prediction for its branching fraction~\cite{Parrott:2022zte} (also see \cite{Becirevic:2023aov}):
\beq
\label{eq:SMBR}
{\cal B}_\text{SM}  (B^+ \to K^+ \nu\nu) = (5.58 \pm 0.37) \times 10^{-6}
\,.
\eeq
The Belle II collaboration recently reported ~\cite{Belle-II:2023esi} a measurement of this branching fraction by employing the novel Inclusive Tagging Analysis (ITA) method \cite{Belle:2019iji,Belle-II:2021rof,Belle-II:2023esi}:
\beq
{\cal B} (B^+ \to K^+ \nu\nu) = (2.7 \pm 0.5 \text{(stat)} \pm 0.5 \text{(syst)}) \times 10^{-5}
\,,\label{eq:Belle-II_Br}
\eeq
which is obtained by scaling and fitting the normalization of a SM-like event distribution.
Ref.~\cite{Belle-II:2023esi} also combines this with results from the well-established Hadronic Tagging Analysis (HTA) method, and obtains ${\cal B} (B^+ \to K^+ \nu\nu) = (2.3 \pm 0.5 \text{(stat)} ^{+0.5}_{-0.4}\text{(syst)}) \times 10^{-5}$. This is higher than the SM prediction by about $2.7 \sigma$, prompting a burst of theoretical activities to look for NP via this mode~\cite{Abdughani:2023dlr, Datta:2022zng, Datta:2023iln, Berezhnoy:2023rxx, Felkl:2023ayn, Bause:2023mfe, Dreiner:2023cms, Allwicher:2023syp, Athron:2023hmz, He:2023bnk, Crivellin:2022obd, Altmannshofer:2023hkn, McKeen:2023uzo}.

Assuming the excess is due to NP, there are two simple interpretations of the ``$\nu\nu$'', which is undetected experimentally. 
One (the ``2-body scenario'') is that ``$\nu\nu$'' is a single new particle, $X$, which is invisible or decays invisibly experimentally \cite{Altmannshofer:2023hkn, McKeen:2023uzo}.
Another (the ``3-body scenario'') is that it is a pair of particles $\chi_1\chi_2$, where each $\chi_i$ may be either a new particle or a SM \mbox{(anti-)}neutrino (e.g.~\cite{Bause:2023mfe, He:2023bnk, Abdughani:2023dlr, Datta:2022zng}). 
The 3-body scenario can be further categorized into three sub-scenarios depending on whether the $b \to s$ transition happens via a scalar, vector, or tensor operator.
The question, therefore, is which scenario seems to be most favored by data on statistical grounds.

In this work, we perform likelihood analyses mainly using the binned ITA data of Ref.~\cite{Belle-II:2023esi}, and identify the preferred ranges of the mass and branching fraction in each scenario.
Then, comparing the best-fit points between different scenarios, we find that the vector-current 3-body scenario with $m_{\chi_{1,2}} \simeq 0.6\>\text{GeV}$ is most statistically favored, but the 2-body decay with $m_X \simeq 2$\,GeV is also comparable. 
We also find that in the scalar- and tensor-current 3-body scenarios the best-fit branching fractions are a few times larger than the aforementioned value by Belle II\@.
This is because the scalar and tensor event distributions have a very different shape from a SM-like (vector) event distribution assumed in the Belle II analysis.
Finally, we also explore further discrimination between scenarios by including the past BaBar data.

\section{Testing new physics with Belle~II} \label{sec:methods}

Given the latest ITA, Belle~II reports the results in terms of a variable ($\qrecsq$) that is only loosely related to $q^2 \equiv (p_B-p_K)^2$, we carefully analyze distortions in the reported signal shape compared to the true $q^2$ distributions expected in the various scenarios. We will also find it essential to take into account the effects of a very non-uniform signal efficiency. We further combine the results with past measurements to narrow down the possibilities.

\begin{figure}[t]
\begin{center}
\includegraphics[width=0.45\textwidth]{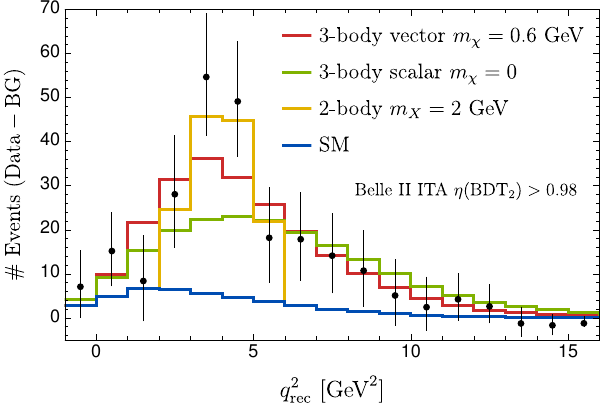}
\vspace{-10pt}
\caption{\label{fig:BelleIIshape} Number of $B^+\to K^+\nu\nu$ events at Belle II~\cite{Belle-II:2023esi} (black dots, with error bars shown) after background subtraction, as a function $q^2_\text{rec}$, using inclusive tagging in the $\eta(\text{BDT}_2)>0.98$ signal region. In blue is shown the SM distribution~\cite{Parrott:2022zte,Becirevic:2023aov}. The red line shows the predicted distribution of events for a 3-body decay $B^+\to K^+\chi\chi$ by a vector current, with $m_\chi=0.6$~GeV and $\mathcal{B}(B^+\to K^+\chi\chi)=3.2\times 10^{-5}$ in addition to the SM. The green line shows $B^+\to K^+\nu_L\chi$ via a scalar current for $m_\chi=0$ and $\mathcal{B}(B^+\to K^+\chi\chi)=7.3\times 10^{-5}$. The yellow line shows the distribution for the 2-body decay $B^+\to K^+ X$ with $\mathcal{B}(B^+\to K^+ X)=0.7\times 10^{-5}$ and $m_X=2$~GeV.
}
\end{center}
\end{figure}

Since the 4-momentum of the tagged $B$ meson is not reconstructed in ITA, the  missing mass-squared, $q^2 = (p_B-p_K)^2$, is approximated by $\qrec^2$:
\begin{align}
q^2_{\text{rec}} &\equiv E_{B}^2 + m_K^2 - 2 E_{B} E_{K} \nonumber
\\
&=q^2 +(E_{B}^2-m_B^2)- 2 \ve{p}_K(q^2) \!\cdot\! \ve{p}_B, 
\end{align}
where $E_{B,K}$ and $\ve{p}_{B,K}$ are the energies and momenta of the observed $B$ and $K$ in the $B\bar{B}$ center-of-mass frame. 
The resolution in $E_K$ measurement is $O(1)$\%, leading to negligible smearing in $\qrecsq$. Given the bin size of $1\text{ GeV}^2$ chosen in Ref.~\cite{Belle-II:2023esi}, the smearing in $\qrecsq$ almost entirely stems from the absence of the $\ve{p}_K(q^2) \cdot \ve{p}_B$ term in the inference of $\qrecsq$.
By numerical simulation, we find that events with a given value of $q^2$ get distributed almost uniformly in $\qrecsq$ as follows: 
\begin{align}
 &f^{{\rm rec}}_{q^2}(\qrecsq) = 
  \begin{cases}(\Delta_+ -\Delta_-)^{-1}& \mbox{if } \Delta_- < \qrecsq -q^2 < \Delta_+  \\ 0 & \mbox{otherwise} \end{cases}, \\
   &\Delta_\pm =(E_{B}^2-m_B^2) \pm 2\left(1 \pm \frac{|\ve{p}_B|}{E_B}\right) |\ve{p}^*_K(q^2)| |\ve{p}_B| \label{eq:fsmear}
\end{align}
where $\ve{p}^*_K(q^2)$ is the momentum of the $K$ in the rest frame of the $B$ meson, $|\ve{p}_B|=0.33$\,GeV, and $E_B=5.29$\,GeV.   
For example, $B \to K X$ events with $m_X = 2$~GeV would be spread over $2.5\,{\rm GeV^2}<\qrecsq<5.5\,{\rm GeV^2}$ as shown in Fig.~\ref{fig:BelleIIshape}. Similarly, the SM distribution in Fig.~\ref{fig:BelleIIshape} is obtained by applying the momentum spreading to a vector current for massless fermions with a total branching ratio as given in Eq.~\eqref{eq:SMBR}. The signal yield of $B^+ \to K^+ \nu\nu$ is extracted from the top-right panel in Fig.~18 of Ref.~\cite{Belle-II:2023esi} by subtracting the $B\bar B$ and continuum backgrounds, which is shown as black dots in Fig.~\ref{fig:BelleIIshape}.

\begin{figure}[t]
    \centering
    \includegraphics[width=0.45\textwidth]{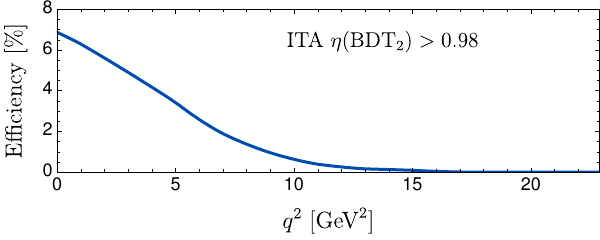}
    \caption{\label{fig:belleiieff}
    Inferred efficiency in the $\eta(\text{BDT}_2)>0.98$ signal region with the inclusive tagging at Belle II.}
\end{figure}

\begin{figure*}[t]
\begin{center}
\includegraphics[width=0.37\textwidth]{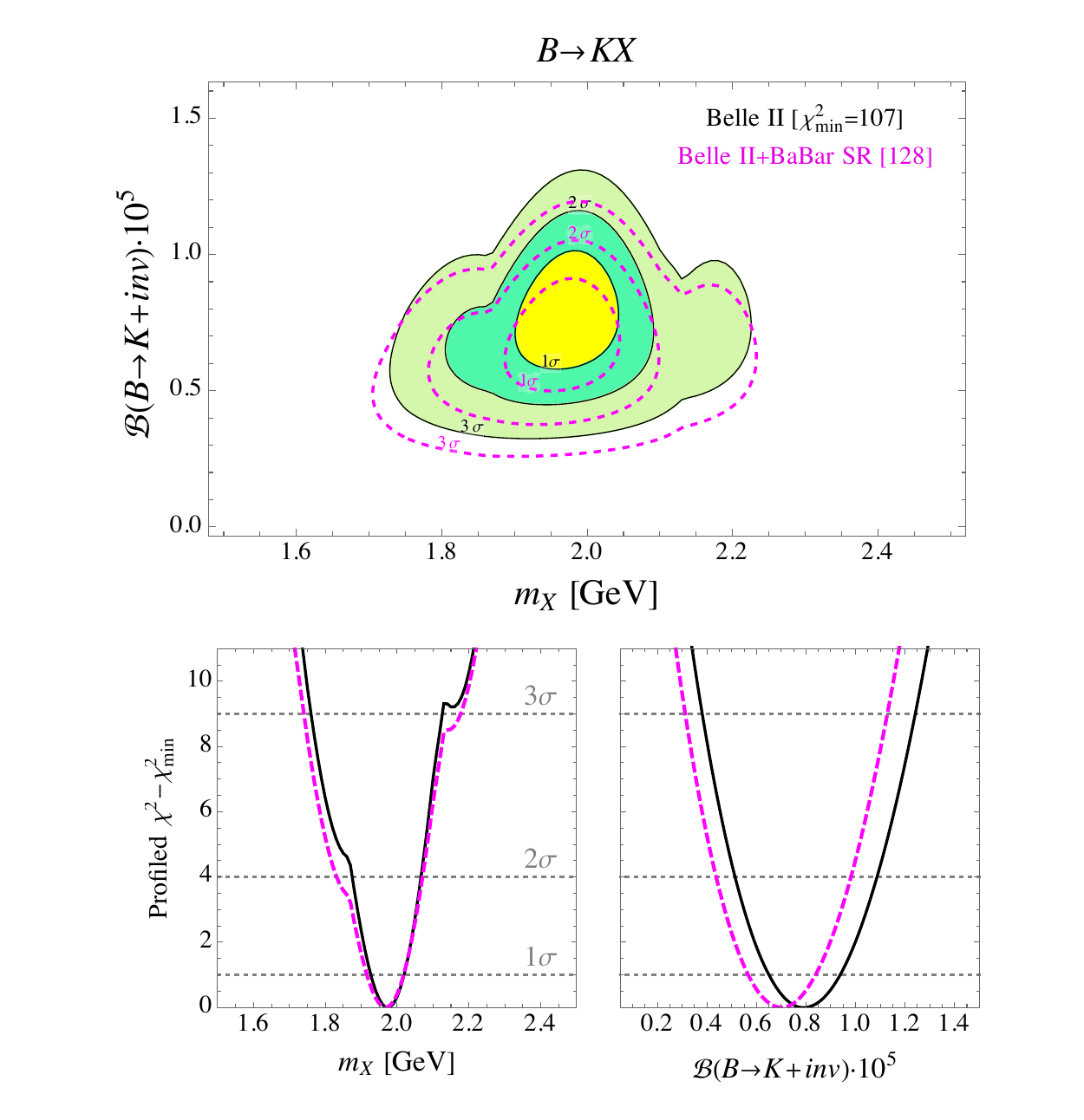}\hspace{-4em}~
\includegraphics[width=0.37\textwidth]{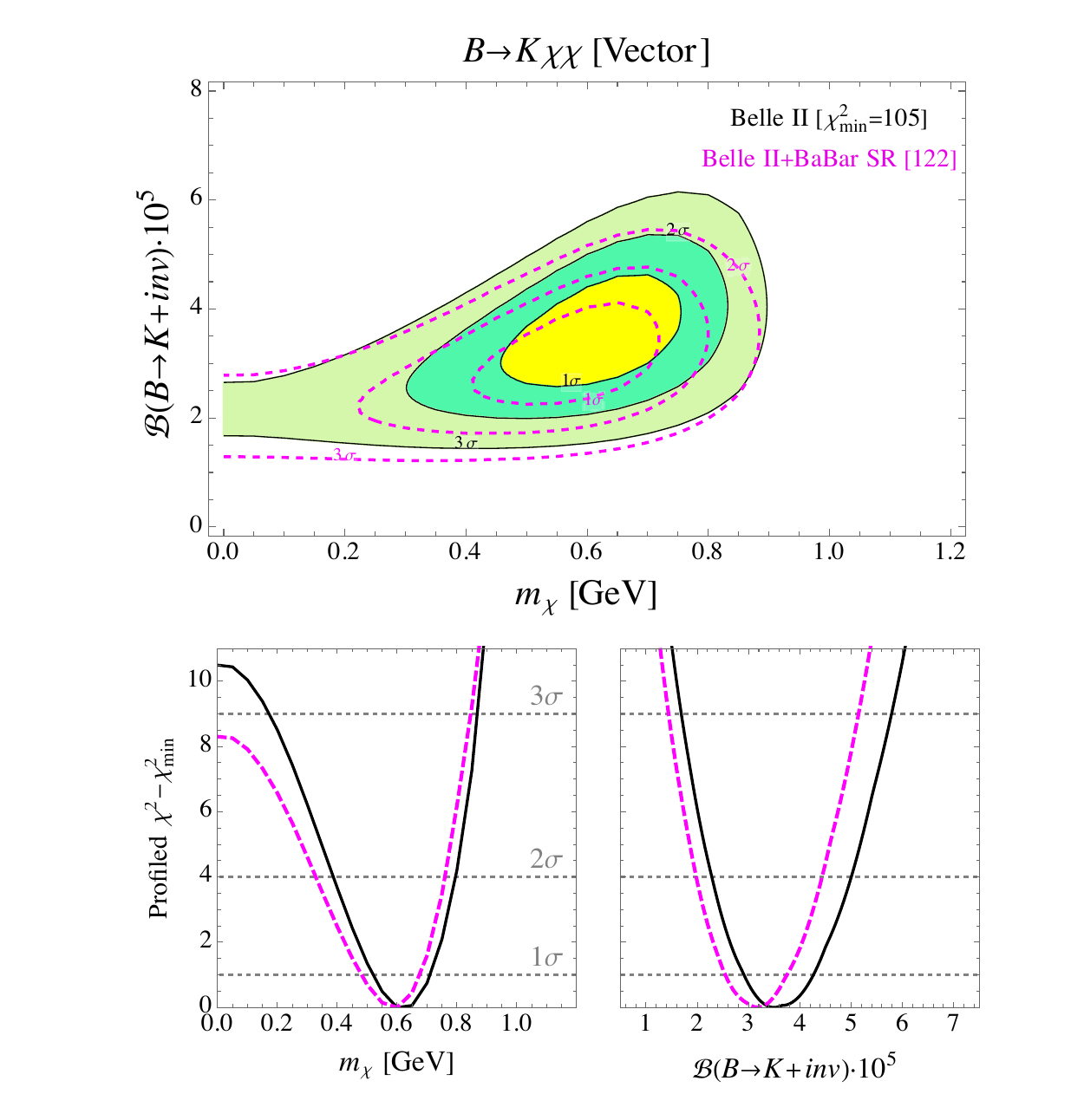}\hspace{-4em}~
\includegraphics[width=0.37\textwidth]{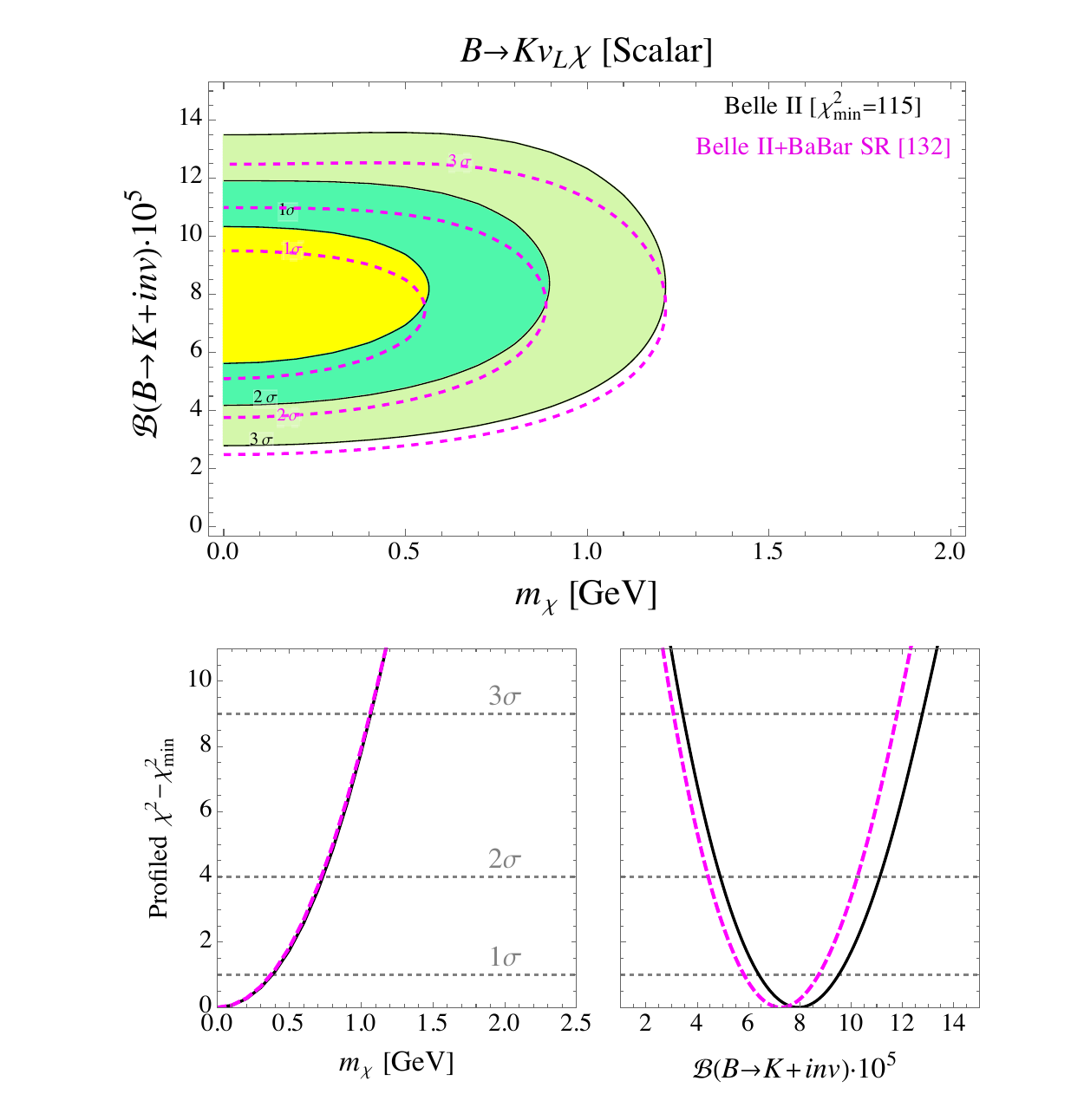}~
\caption{\label{fig:chi2min}The upper panel shows 2D plots of $\chi^2 - \chi^2_{\min}$ in the ${\cal B}$-${m_{\rm NP}}$ plane for different scenarios. For each plot in the upper panel, the likelihoods profiling one variable  (along the ${\cal B}$ direction and along the $m_\text{NP}$ direction) are provided in the lower panel. The three columns correspond to different NP scenarios: on the left panel is the 2-body scenario $B^+ \to K^+ X$, for which $m_{\rm NP} = m_X$, the center panel is for the 3-body decay $B^+ \to K^+ \chi \chi,\, m_{\rm NP} = m_{\chi}$ mediated by a vector current, while the right column corresponds to the 3-body decay $B^+ \to K^+ \bar{\chi}_R\nu_L , m_{\rm NP} = m_{\chi}$, where $\nu_L$ is a SM neutrino.}
\end{center} 
\end{figure*}

The signal efficiency, $\epsilon(q^2)$,  is another essential input to process the various scenarios. Although two efficiencies are reported in the latest Belle II analysis, they are not applicable to the samples with the highest ``BDT$_2$'' cut, BDT$_2>0.98$, where the excess from the SM background is visible. 
Therefore we deduce the efficiencies to reproduce the shape of the reported $B\to K\nu\nu$ distribution. We tune the $q^2$-dependent efficiency such that the vector-current mediated distribution with the best-fit normalization of the ITA analysis, ${\cal B}=2.7\times 10^{-5}$, is multiplied by $q^2$-dependent efficiency and binned with the $\qrecsq$ smearing, and then the resultant distribution is matched with the one reported in Fig.~18 of Ref.~\cite{Belle-II:2023esi}. 
 The resulting efficiency, shown in Fig.~\ref{fig:belleiieff} after interpolation in $q^2$, is similar in shape to the one with BDT$_2>0.95$ reported in the earlier Belle~II analysis~\cite{Belle-II:2021rof}.

For the statistical combination, we construct the binned likelihood, $L =\prod_i f_{\rm P} (N_i^{{\rm obs}}; N_i^{\rm ex} )$, using Poisson statistics for the $i$th bin 
 $q^2_{i} \leq \qrecsq< q^2_{i}+1\,{\rm GeV^2} $. The expectation $N_i^{\rm ex} =N_i^{\rm BG}+N_i^{K\nu\nu,\rm SM}+N_i^{\rm NP}$ includes the background and the SM $B\to K\nu\nu$ with as well as the yield of new physics,
\begin{align}
N_i^{\rm NP}  = N_{B^{\pm}} \int_{q^2_i}^{q^2_{i+1}}\hspace{-3ex} \D x 
 \int \D q^2 \, f^{{\rm rec}}_{q^2}(x) \, \epsilon(q^2) \frac{\D {{\cal B}_{\rm NP}}}{\D q^2} \,,
\end{align}
where $N_{B^\pm} = 3.99\times 10^8$ at Belle~II\@. 
We obtain the log-likelihood, $-2\ln L$, which is denoted as  $\chi^2$ for simplicity. This method accounts for only the statistical uncertainty. 
As a check, we obtain ${\cal B}=(2.6\pm 0.4\text{(stat)})\times 10^{-5}$ for  $B^+\to K^+\nu\nu$ for ITA, which is reasonably consistent with ${\cal B}=(2.7\pm 0.5\text{(stat)})\times 10^{-5}$ reported by Belle~II.

\begin{figure}
    \centering
    \includegraphics[width=0.4\textwidth]{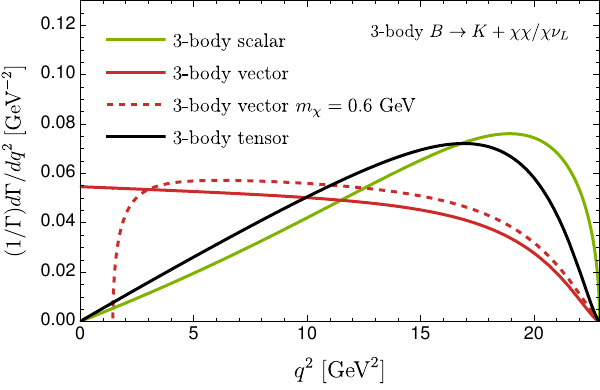}
    \caption{\label{fig:svtDistribution}
    Normalized distributions for  $B^+\to K^+\chi_1\chi_2$ via scalar, vector, and tensor operators as functions of $q^2$ for massless $\chi_{1,2}$ (solid lines)  as well as massive $\chi_{1,2}$ (dotted line)  with an equal mass of $0.6 \, \text{GeV}$. The distributions do not take experimental efficiency into account.}
\end{figure}


\section{Testing 2-body scenario}\label{sec:2-body}
First, we calculate the log-likelihood $\chi^2$, based on the 2-body decay $B^+ \to K^+ X$ scenario. We find the minimum $\chi^2_{\rm min}$ and obtain the preferred parameter space by evaluating $\chi^2-\chi^2_{\rm min}$. Since there is a significant mass preference, we perform the 2D fit in the $m_X$-$\cal B$ space, and the contours with 1, 2, and 3$\sigma$ are shown in the upper-left pane in Fig.~\ref{fig:chi2min}.

Although our method correctly accounts for the statistical uncertainty, the systematic uncertainty can be comparable based on the Belle~II report. 
We expect that including systematic uncertainty lowers the $\chi^2$ by a factor of two, resulting in the allowed region being enlarged by $\sim 40\%$, but the tendency of the preferred parameters would not be affected.

To extract the preferred mass, we use the profile log-likelihood: finding the best $\chi^2$ for each mass by profiling $\cal B$, which is shown in the bottom left panel. We find a narrow range of mass, {$m_X=1.97\pm 0.05$(stat)\,GeV}. The example of distribution in the ITA is shown in Fig.~\ref{fig:BelleIIshape}. 
A similar method is applied for the branching fraction, leading to {${\cal B}=\left(0.79\substack{+0.16 \\ -0.13}{\rm (stat)}\right)\times 10^{-5}$}. 
This is in accordance with a recent study of the 2-body  scenario~\cite{Altmannshofer:2023hkn}, but it does not agree with the other~\cite{McKeen:2023uzo}. We wish to emphasize that the expected number of events in the given $\qrecsq$ binning is calculated following Eq.~\eqref{eq:fsmear} so that we appropriately fit the signal with the data of the ITA.

\section{Testing 3-body scenarios}\label{sec:3-body}
The same method is applied to the 3-body case $B^+ \to K^+ \chi_1 \chi_2$ where we have multiple scenarios to consider.  We assume that $\chi_{1,2}$ are fermionic in this work. However, some operators that contribute to $B \to K$ decays also contribute to $B\to K^*$ decays. Belle-II has recently obtained an upper limit on the branching fraction ${\cal B} (B^0\to K^* \nu\nu)< 1.8 \times 10^{-5}$ \cite{Belle:2017oht}, but a measurement of the branching fraction has not been obtained yet. We consider operators that do not have significant effects on $B\to K^*$: 
\begin{enumerate}

    \item Scalar operator: the interaction is given by 
    \beq {\cal L} \supset {1 \over \Lambda_S^2} (\bar{b} s) (\bar{\chi}_1 \chi_2), \label{eq:scalarop} 
    \eeq
   where $\Lambda_S$ is a heavy scale. We strictly distinguish this case from the pseudo-scalar case with $\bar{b} \gfive s$ as that would contribute only to $B \to K^*$ and not to $B \to K$, while the scalar case contributes only to $B \to K$ and not to $B \to K^*$.
    In our work, we consider the case where $\chi_1 = \nu_L$, which is massless for our purpose, and $\chi_2$ is a new fermion, possibly massive. Note that in Ref.~\cite{Deppisch:2020oyx} it was shown that a scalar current distribution for the related kaon decay mode $K^+\to\pi^+\nu\nu$ implies lepton number violation (LNV) for SM-invariant operators, since such a current can only be generated via $\Delta L= 2$ odd mass-dimension operators. We consider the possibility of $\chi_2$ being the right-handed neutrino $N$, for which case there exists non-LNV operators at dimension-6 which generate a scalar current, e.g.\ the operator $\mathcal{O}_{LNQd}=\epsilon_{ij}\overline{L^{i}}N \overline{Q^j}d$~\cite{Liao:2016qyd}.
    
    \item Vector operator: the interaction is given by
    \beq {\cal L} \supset {1 \over \Lambda_V^2} (\bar{b}\gmu  s) (\bar{\chi}_1 \gamma_{\mu} \chi_{2})
    \,,\label{eq:vecop}
    \eeq
    where $\Lambda_V$ is a heavy scale. $\chi_1$ and $\chi_2$ may be a pair of SM $\nu_L\nu_L$ or that of new (massive) neutral fermions.
    The quark part of this operator is a pure vector, which is crucially different from a SM-like $V$$-$$A$ current. As shown in Fig.~2 of \cite{Bause:2023mfe}, an additional $V$$-$$A$ contribution that fits the $B\to K$ excess would be severely excluded by the absence of a corresponding excess in $B\to K^*$, while the pure vector case has the least impact on $B\to K^*$ and is still allowed by data. Therefore, for our interest in identifying the best scenarios, we do not consider any axial-vector component in the quark bilinear.
    
    \item Tensor operator: the interaction is given by
    \beq
    {\cal L} \supset {1 \over \Lambda_T^2} (\bar{b} \sigma^{\mu \nu} s)
   (\bar{\chi}_1 \sigma_{\mu\nu} \chi_2)\,,\label{eq:tensorop}
    \eeq
    where $\Lambda_T$ is a heavy scale. As in the scalar case above, 
    we consider $\chi_1=\nu_L$ and $\chi_2$ to be a new neutral fermion.
    
\end{enumerate}
In each of these three cases, we compute the $q^2$ spectrum of the decay rate $\D \Gamma / \D q^2$, and the distributions are given in Fig.~\ref{fig:svtDistribution}. 
The relevant matrix elements and form factors for the $B\to K$ decay are given in \cite{Isgur:1990kf,Ball:2004ye} and also
in Appendix ~\ref{app:formfactors}.
In the ITA, the distributions are modified due to the efficiency and smeared by $\qrecsq$. Two representative cases are shown in Fig.~\ref{fig:BelleIIshape}.


We find the favored parameter space of the 3-body scenarios using the same likelihood analysis as we perform for the 2-body one, and the results for the cases of the vector and scalar operators are shown in Fig.~\ref{fig:chi2min}. 
For the vector case, the fit is best at $m_{\chi}= 0.62\substack{+0.10 \\ -0.09} \, \text{GeV}$ and ${\cal B}=(3.5\substack{+0.8 \\ -0.6}(\text{stat})) \times 10^{-5}$ (see the middle panel), which is close to the measurement made at Belle~II. Since this massive $\chi$ be can be stable, it is a candidate of dark matter.

For the scalar operator, the best-fit mass is $m_{\chi} = 0$, while the branching fraction is preferred to be high, {$(7.9 \substack{+1.6 \\ -1.5}(\text{stat}) \times 10^{-5})$} (see the right panel). 
This is due to that the $q^2$-spectrum peaking at high $q^2 \sim 20 \  \text{GeV}^2$, as seen in Fig.~\ref{fig:svtDistribution}, is multiplied by the efficiency which drops at high $q^2$, see Fig.~\ref{fig:belleiieff}.  
Therefore, most of the signal events are discarded by the ITA, which statistically favors a higher branching fraction to explain the excess.

For the tensor case, since the $q^2$-distribution is very similar to the scalar case, as seen in Fig.~\ref{fig:svtDistribution}, the fitted results are also similar, and the preferred branching fraction is ${\cal B}=(6.4\pm 1.2(\text{stat})) \times 10^{-5}$. 
We do not show the tensor case in Fig.~\ref{fig:chi2min} for the sake of brevity, but we quote $\chi^2_{\text{min}}$ for the fit to the decay $B \to K \chi \nu$ in Table~\ref{tab:chi2_list}.

For the best-fit points in Fig.~\ref{fig:chi2min}, we find $\Lambda_V\approx 6.5$~TeV for the 3-body vector current and $\Lambda_S\approx 4.1$~TeV for the 3-body scalar current. The 2-body decay occurs at tree-level, for which we find an effective coupling $\lambda_{X}\approx 3\times 10^{-4}$ for $\mathcal{L}\supset -\lambda_{X}\bar{b}sX$.

As already addressed, our analysis does not include systematic uncertainty. The systematic uncertainty would enlarge the allowed parameter space by about $40\%$ unless the uncertainty has a strong $q^2$ dependence. 


Based on the values $\chi^2_{\min}$ of different scenarios, shown in Table~\ref{tab:chi2_list}, we can distinguish the preferred ones out of them. Testing one hypothesis $H_1$ against  the null hypothesis $H_0$ can be evaluated by \cite{Workman:2022ynf}
\begin{align}
   \alpha= \frac{\max L(H_1)}{\max L(H_0)}=\exp\left(-[\chi_{\rm min}^2(H_1)-\chi_{\rm min}^2(H_0)]/{2} \right)\, .
\end{align}
Within the Belle~II ITA, we find that the best scenario is the 3-body decay with vector-operator and $m_\chi= 0.62$\,GeV, and the best 2-body scenario is also competitive. However, the scalar-current 3-body scenario is disfavored, and adding mass to $\chi$ makes it even worse. A SM-like shape, i.e. the vector-current 3-body scenario with massless $\chi$, is also disfavored.
Even if unaccounted uncertainties, such as the systematics, lower $\chi^2$, say by a factor of two, the scalar and vector operators with massless $\chi$ would still have $\alpha<10\%$ against the best-fit scenario.

\begin{table}[t!]
    \centering
    \setlength{\extrarowheight}{6pt}
    \begin{tabular}{|m{1mm} m{27mm} | m{1mm} m{7mm} m{7mm}  m{7mm}  m{7mm}  m{7mm} | m{8mm}|}
    \hline
    \specialrule{2pt}{2pt}{0pt}
    & $\chi^2_{\rm min}-100$ & & 2b & V & V$'$ &  S &T & SM \\[2mm]
   \specialrule{2pt}{0pt}{0pt}
   & Belle II & &{\bf 6.8}  &15.2& {\bf 4.7}&15.1 &11.9 &44.6 \\[2mm]
   \specialrule{1pt}{0pt}{0pt}
   & + BaBar~SR & & 27.6 & 30.4 &{\bf 22.1}  &31.8  & 29.8 & 61.0  \\[2mm]
   \hline
   & + BaBar~$s_B<0.8$ & & {\bf 73.3} &78.8 & {\bf 72.9}&{90.2}  & 86.9 & 106.7 \\[2mm]
      \specialrule{2pt}{0pt}{2pt}
      \hline
    \end{tabular}
    \caption{The likelihood minima $\chi^2_{\text{min}}$ for 2-body scenario (2b), and several 3-body scenarios: two vector cases with $m_\chi=0$ (V) and $m_\chi\simeq 0.6 \, \text{GeV}$ (V$'$);  the scalar (S)  and tensor (T) cases. 
    The rows correspond to the choice of analyzed datasets: first only the Belle~II ITA data \cite{Belle-II:2023esi}, then adding the BaBar data~\cite{BaBar:2013npw} within the signal region (SR), and finally including outside the SR, $q^2 < 0.8 m_B^2$.  
    The bold ones indicate the best-fit or competitive scenarios with $\alpha(H_0=\rm V')>10\%$.  The $\chi^2$ values for the SM  are also shown.
    Only statistical uncertainties are considered.
    }
    \label{tab:chi2_list}
\end{table}

\paragraph{\textbf{Combining with other measurements.} ---}\label{sec:combine}

\begin{figure}[t]
\begin{center}
\includegraphics[width=0.45\textwidth]{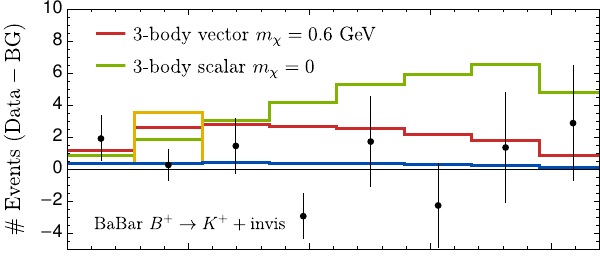}\\
\vspace{-5pt}
\includegraphics[width=0.45\textwidth]{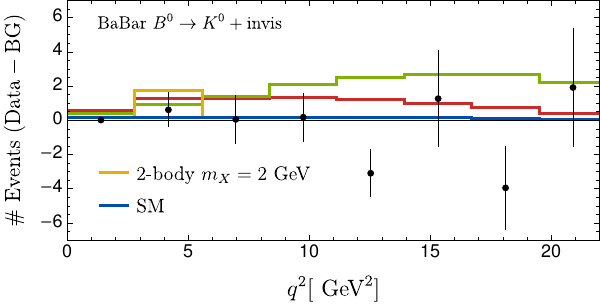}
\caption{ \label{fig:babarshape}
Number of $B\to K\nu\nu$ events (black, error bars shown) for $471\times 10^6$ $B\bar B$ pairs at BaBar~\cite{BaBar:2013npw} with background subtracted, as a function of $q^2$. The colored lines show the same benchmark scenarios as Fig.~\ref{fig:BelleIIshape}.
}
\end{center}
\end{figure}

Having considered the Belle II ITA, we now combine the analysis with past BaBar measurements of $B \to h \nu\nu$ with $h = K^0, K^+ $ based on the conventional hadronic tagging \cite{BaBar:2013npw}.
The shape with the BaBar data (after the background is subtracted) with several new physics scenarios in Fig.~\ref{fig:babarshape}. 
A major difference from the latest Belle II ITA is that the analysis uses $q^2$ rather than $\qrecsq$, and the resolution of $q^2$ is significantly better than the adopted bin size. Therefore, we can ignore smearing effects.

For their signal region  $s_B\equiv q^2/m_B^2<0.3$, we calculate the log-likelihood $\chi^2$ for this data, which we then combine with the result of Belle~II ITA. The combined fits are shown as magenta dashed lines in Fig.~\ref{fig:chi2min}, and the values of $\chi^2_{\min}$ are in Table~\ref{tab:chi2_list}. While the preferred parameter regions barely change in all scenarios,  the difference of $\chi^2_{\min}$ between the 2-body scenario and the best 3-body scenario increases from 2 to 5 because the signal of 2-body decay should be localized in the second bin where no excess is seen.

Although $s_B >0.3~(q^2\gtrsim 8.4\,{\rm GeV}^2)$ is outside the signal region and requires more careful treatment of the uncertainties, it is interesting to examine this region for further discrimination. For some 3-body scenarios with the scalar and tensor operators, we expect an excess in high $q^2$ because the efficiency in this BaBar study stays flat until $q^2\sim 20\,{\rm GeV}^2$ while the efficiency of the Belle~II ITA is very small for $q^2> 10\,{\rm GeV}^2$. Thus, increasing the branching fraction to fit the Belle II ITA would be in tension with $B^+ \to K^+\nu\nu$ result at BaBar, as seen in Fig.~\ref{fig:babarshape}. For this reason, the $\chi^2_{\min}$ values of these scenarios are significantly worsened, see Table~\ref{tab:chi2_list}. The best scenario is still the 3-body decay with massive $\chi$ via the vector operator, but the 2-body one is comparable.

\begin{figure}
    \centering
    \includegraphics[width=0.45\textwidth]{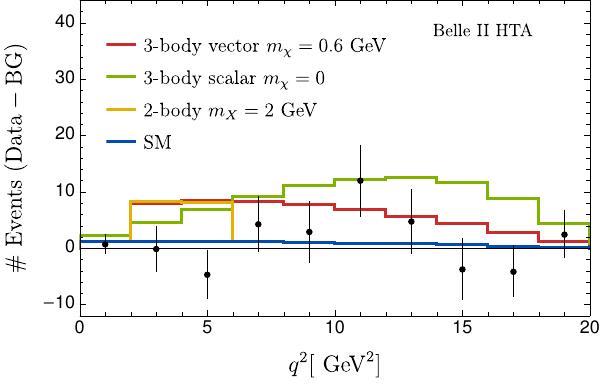}
    \caption{\label{fig:HTA} 
    Number of $B^+\to K^+\nu\nu$ events at Belle II~\cite{Belle-II:2023esi} (black dots, with error bars shown) after background subtraction using hadronic tagging, as a function of $q^2$. The colored lines show the same benchmark scenarios as in Fig.~\ref{fig:BelleIIshape}. 
    }
\end{figure}

Other data sets that may be relevant for us, but are not included, are the following:

\begin{enumerate}
    \item In \cite{Belle-II:2023esi}, Belle~II also reported the $q^2$ result in the HTA, see Fig.~\ref{fig:HTA}. Since the data has a small overlap with the ITA dataset, we are unable to combine them. However, we see the trend that the third bin seems to have a tension with the 2-body scenario with $m_X=2$\,GeV. 
    
    \item Belle results via hadronic \cite{Belle:2013tnz} and semileptonic \cite{Belle:2017oht} tagging: the number of events is measured in bins of $E_{\rm ECL}$, which is the residual energy deposited in the electromagnetic calorimeter (ECL). Since $E_{\rm ECL}$ is not related to $q^2$, we are unable to include this data for testing the various scenarios. 

    \item BaBar results via semileptonic tagging \cite{BaBar:2010oqg}: The binning is done in bins of $|p_K|$, the magnitude of the momentum of the $K^+$ in the center-of-mass frame. While $|p_K|$ has a one-to-one correspondence to $\qrecsq$, the final results are reported as an average of 20 different BDT analyses. Due to this unconventional statistical treatment, we cannot safely add this data to our analysis. 
\end{enumerate}

\section{Conclusion}\label{sec:conclusion}
In this work, we perform a  binned-likelihood analysis of different NP scenarios mainly based on the recent Belle II ITA data on the decay $B^+ \to K^+ \nu\nu$. The NP scenarios we consider are: (a) 2-body scenario $B \to K X$, and (b) the 3-body scenario $B \to K \chi_1 \chi_2$. For (b) we consider several operators (scalar, vector, and tensor) as well as several masses for the new particles. 

We find it crucial to account for the fact that the Belle~II ITA data is binned not in the momentum transfer $q^2$ but in $\qrec^2$, and for the non-uniform efficiency. 
We augment our analysis with past BaBar data. Our results are listed in Table~\ref{tab:chi2_list} and in Fig.~\ref{fig:chi2min}.

As seen in Table~\ref{tab:chi2_list}, the best fit NP scenario is a 3-body decay of the $B^+$ via a vector current into $K^+$ and a $\chi \chi$ pair, where $m_{\chi} \simeq 0.6 \, \text{GeV}$. This $\chi$ can be stable and hence a possible dark matter candidate. Table \ref{tab:chi2_list} also tells us that the 2-body scenario is not much worse, and so it still remains competitive. 

The right panel of Fig.~\ref{fig:chi2min}, shows the preferred parameter region in the scalar-current 3-body scenario. Massless $\chi$ fits best, for a branching fraction of ${\cal B} (B^+ \to K \chi_R \nu_L) \sim 8 \times 10^{-5}$. This is about a factor of 3 higher than the value in Eq.~(\ref{eq:Belle-II_Br}). This is an especially clear demonstration of the importance of considering the non-uniform signal efficiency in inferring the branching fraction of NP from the reported excess at Belle~II.


\begin{acknowledgements}
We thank Eldar Ganiev for useful discussions. This work is supported in part by the US Department of Energy grant DE-SC0010102. The work of K.F., T.O., and K.T.~is supported in part by JSPS Grant-in-Aid for Scientific Research (Grant No.\,21H01086).
K.F.~acknowledges support from the JSPS KAKENHI grant JP22K21350.

\end{acknowledgements}





\appendix

%
%
%
%
%
%
%

\section{Hadronic Matrix Elements and Form factors for $B \to K$}
\label{app:formfactors}

The hadronic matrix elements of some of the operators relevant to $B \to K$ decays are given below \cite{Isgur:1990kf,Ball:2004ye}:
   \begin{eqnarray}
       \mel{\bar{b}s}{K}{B} &=& {m_B^2 -m_K ^2 \over m_s -m_b} f_0 (q^2) \,,  \\
        \mel{\bar{b} \gmu s}{K}{B} &=& f_+ (q^2) (p_B + p_K)^{\mu}  \\
        && + \bigl[ f_0 (q^2) -f_+(q^2) \bigr] {m_B^2 -m_K ^2 \over q^2} q^{\mu} \,, \nonumber\\
         \mel{\bar{b} \sigma^{\mu \nu} s}{K}{B} &=& i {f_T (q^2) \over m_B + m_K} \bigl[(p_B + p_K)^{\mu} q^{\nu} \nonumber\\
         &&\hspace{13ex} -(p_B + p_K)^{\nu} q^{\mu} \bigr] \,,
   \end{eqnarray}
       where $q^\mu = p_B^\mu -p_K^\mu$ and the form factors $f_0, f_+$ and $f_T$ can be given approximate analytical expressions via lattice methods. Here we quote the results obtained in Ref.~\cite{Ball:2004ye}:
       \begin{eqnarray}
           f_0 (q^2) &=& {r_1 ^P + r_2 ^P\over 1 -q^2/m_{\text{fit}}^2} \,,\\
           f_+(q^2) &=& {r_1^P \over 1 - q^2/M_B^2} + {r_2^P \over (1- q^2/M_B^2)^2} \,,\\
           f_T (q^2) &=& {r_1^T \over 1 - q^2/M_B^2} + {r_2^T \over (1- q^2/M_B^2)^2} \,,
       \end{eqnarray}
       where $r_1^P = 0.162$, $r_2^P = 0.173$, $r_1^T = 0.161$, $r_2^T = 0.198$, $M_B = 5.41\;\text{GeV}$ and $m^2_{\text{fit}} = 37.46\;\text{GeV}^2 $ are fit parameters in the lattice.

Note that using isospin symmetry, we claim that the form factors for the charged $B \to K$ modes are essentially identical. Using these relations we can find expressions for the branching fraction for NP contributions to $B^+\to K^+ +$ invisible. For example, the branching fraction for an additional vector current contribution to $B^+\to K^+\nu\nu$ is given by
   \begin{equation}
   \label{eq:NPcontribution}
   {\cal B}_\text{NP}(B^+\to K^+\nu\nu)\times \Gamma_{B^+}^\text{tot} = \int_{0}^{(q^2)^+} \hspace{-4.5ex} \D q^2\int_{t^-}^{t^+} \hspace{-1.5ex} \D t \, \frac{\D \Gamma_V}{\D q^2 \D t}
   \end{equation}
   where $(q^2)^+=(p_{B^+} - p_{K^+})^2$ and  
   \begin{equation}
   \begin{aligned}
   t^\pm &= \frac{(m_{B^+}^2 -m_{K^+}^2)^2}{4q^2} \\
   &\qquad -\frac{1}{4q^2}\left(\lambda^{1/2}({q^2,m_{B^+}^2,m_{K^+}^2})\mp q^2\right)^2\, ,
   \end{aligned}
   \end{equation}
   where $\lambda(a,b,c)$ is the Källén function, given by:
   \beq \lambda (x,y,z) \equiv x^2 + y^2 + z^2 - 2xy -2yz -2xz. \eeq
   
   Furthermore, we have
   \begin{equation}
   \begin{aligned}
   \frac{\D \Gamma_V}{\D q^2 \D t} =& \frac{1}{(2\pi)^3}\frac{1}{32m_{B^+}^3}\left(\frac{1}{\Lambda_V^2}\right)^2\\
   &\times |\mel{\bar{b} \gmu s}{K}{B} \mel{\bar{\nu} \gamma_\mu \nu}{\nu\bar\nu}{0}|^2\, ,
   \end{aligned}
   \end{equation}
   where $\Gamma_{B^+}^\text{tot}\approx 4.0\times 10^{-13}$~GeV is the total width of the $B^+$.

\bibliographystyle{apsrev4-1}
\bibliography{mybib}

\end{document}